# Performance Comparison of Graphene Nanoribbon FETs with Schottky Contacts and Doped Reservoirs

Youngki Yoon[1,a], Gianluca Fiori[2,b], Seokmin Hong[1], Giuseppe Iannaccone[2], and Jing Guo[1]

[1] Department of Electrical and Computer Engineering, University of Florida, Gainesville, FL 32611, USA

[2] Dipartimento di Ingegneria dell'Informazione: Elettronica, Informatica, Telecomunicazioni, Università di Pisa, Via Caruso 16, I-56122 Pisa, Italy

**ABSTRACT**

We present an atomistic 3D simulation study of the performance of graphene nanoribbon (GNR) Schottky barrier (SB) FETs and transistors with doped reservoirs (MOSFETs) by means of the self-consistent solution of the Poisson and Schrödinger equations within the non-equilibrium Green's function (NEGF) formalism. Ideal MOSFETs show slightly better electrical performance, for both digital and THz applications. The impact of non-idealities on device performance has been investigated, taking into account the presence of single vacancy, edge roughness and ionized impurities along the channel. In general, MOSFETs show more robust characteristics than SBFETs. Edge roughness and single vacancy defect largely affect performance of both device types.

e-mail: [a] ykyoon@ufl.edu, [b] g.fiori@iet.unipi.it

* The first two authors equally contributed to this work.

# I. INTRODUCTION

In the last decade carbon nanostructures have attracted much attention from the device research community because its electrical properties make it very appealing for electronic applications. Carbon nanotubes were demonstrated for the first time by Ijima [1], and from there huge effort has been directed to understand the physical properties of the new material and to exploit its potentials in electronic applications to come after Moore's law and ITRS requirements [2-4]. Carbon atoms not only can be combined in the form of tubes of nanoscale dimensions, but can also be arranged in a stable two-dimensional graphene sheet [5-7]. Electrons in graphene behave as massless fermions and travel through the lattice with long mean free path, as shown by the high mobility [5-7]. Graphene is a zero gap material, with linear dispersion in correspondence of the Fermi energy, which makes it particularly unsuitable for transistor applications. However, energy gap can be induced by means of lateral confinement [8], realized for example by etching the graphene sheet in narrow stripes, so-called graphene nanoribbons (GNRs).

Theoretical works have shown that GNRs have energy gap which is inversely proportional to their width [9, 10], and due to their reduced dimensions, edge states play an important role, defining non-null energy gap, for all ribbon widths [11, 12]. GNR field-effect transistors (GNRFETs) have been fabricated very recently [13-15]. GNRFETs demonstrated experimentally to date are realized by connecting the channel to metals with Schottky contacts [8, 14], therefore obtaining a Schottky barrier FET (SBFET). In addition, ohmic contacts can in principle be obtained by heavily doping the GNR source and drain extensions, which makes device operation MOSFET-like (therefore, it is referred to as a MOSFET in the subsequent discussion). Since fabrication techniques are at the very first steps, simulations can represent an important tool to evaluate device performance. Semiclassical top-of-the-barrier simulations have been performed [16, 17], while quantum simulations based on a tight-binding approach have followed [18-21] in order to assess device potential. However, due to the embryonic stage of this new field of research, many issues still remain unsolved. It is, for example, not clear how much performance improvement can be obtained by using a MOSFET device structure, as compared to the Schottky contacts counterpart, as well as the extent to which non-idealities can affect device characteristics. State-of-the-art etching techniques are for instance far from atomistic resolution so that edge roughness can play an important role on device performance [22, 23]. In addition, defects or ionized impurities can represent elastic scattering centers, which can greatly degrade the expected fully-ballistic behavior.

In this work, GNR SBFET and MOSFET are numerically studied in order to establish their potential and the performance that can be expected if technological challenges are met. The approach is based on the self-consistent solution of the three-dimensional (3D) Poisson and Schrödinger equations within the non-equilibrium Green's function (NEGF) formalism [24], by means of a real space $p_z$ tight-binding Hamiltonian, in which energy relaxation at the GNR edges is considered. Different types of non-idealities have been investigated. In particular, we have studied the effect of a single vacancy defect, an ionized impurity in the channel, and edge roughness on the device performance. Doped source and drain reservoirs devices show better performance as compared to Schottky GNRFETs. Vacancies and edge roughness can greatly affect device electrical performance, more than ionized impurities actually do.

## II. APPROACH

Device characteristics of GNRFETs are calculated by solving Schrödinger equation using the NEGF formalism [24] self-consistently with 3D Poisson equation [18-21]. A tight-binding Hamiltonian with an atomistic $p_z$ orbital basis set is used to describe atomistic details of the GNR channel. Coherent transport is assumed. Simulated device structures are depicted in Fig. 1. The source and the drain are doped extensions of GNRs in MOSFETs, and metals in SBFETs with Schottky barrier height of $\Phi_{Bn} = \Phi_{Bp} = E_g/2$. Double-gate geometry is used through 1.5 nm SiO$_2$ gate oxide ($\kappa$ = 3.9). For an ideal device simulation, perfectly patterned 15 nm-long $N$ = 12 [9] armchair-edge GNR (A-GNR) is used as a channel material, which has a width of ~1.35 nm and a bandgap of ~0.6 eV. Edge bond relaxation is treated according to *ab-initio* calculation, and a tight-binding parameter of $t_0$ = 2.7 eV is used [11]. Power supply voltage is $V_{DD}$ = 0.5 V. Room temperature (T = 300 K) operation is assumed.

Non-idealities are treated as follows. Lattice vacancies or edge roughness are considered as atomistic defects of the channel GNR, where the existence of carriers is essentially prohibited. These atomistic vacancy or edge roughness can be implemented by breaking the nearest bonds ($t_0 = 0$) in the device channel Hamiltonian matrix of the perfect lattice according to the geometry of the defective lattice. For simplicity, it is assumed that topological structure of GNR is not affected by the defect, which may provide a perturbation to the quantitative results, but the qualitative conclusions of this study will not be changed. An ionized impurity is treated as an external fixed charge, which can play an important role for electrostatic potential of the device. In other words, in the self-consistent iterative loop between the transport equation and the Poisson equation, the input charge into the Poisson equation always includes a fixed external charge as well as the output charge from the Schrödinger equation.

## III. RESULTS

*A. Ideal Structures*

We first present results for an SBFET and a MOSFET under ideal conditions. Figures 2a and 2b depict the transfer characteristics for each device. SBFET shows the typical ambipolar behavior (Fig. 2a), so that, for a fair comparison, a common off current $I_{off} = 10^{-7}$ A is selected and on state is defined at $V_{on} = V_{off} + V_{DD}$. Then the operating voltage ranges are shown by the gray windows in Figs. 2a and 2b for each device. Through the gate work function tuning, $V_{off}$ can be shifted to $V_G = 0$ V ($V_D = V_{DD}$), and the transfer characteristics after the work function engineering are shown in Fig. 2c: the MOSFET has 50 % larger $I_{on}$ (i.e. current for $V_G = V_{DD}$, $V_D = V_{DD}$) and larger transconductance $g_m$ than the SBFET. This observation agrees to a conclusion in a previous literature that the on current of a ballistic SBFET with positive Schottky barrier height is smaller than that of a ballistic MOSFET due to the tunneling barrier at the source end of the channel [25]. MOSFETs can have in addition a significantly larger maximum on-off ratio than SBFETs due to the suppression of ambipolar transport, as shown in Fig. 2d.

Figure 3a shows the output characteristics for $V_G = 0.5$ V: MOSFET shows a better saturation behavior. This is confirmed by the output conductance $g_d$ defined as the derivative of the output characteristic with respect to $V_D$. As can be seen in Fig. 3b, $g_d$ in MOSFET is almost half the value found for SBFET.

We now focus on switching and high-frequency performance of GNR devices. In Fig. 4a the cutoff frequency $f_T$ as a function of the applied gate voltage is shown, and computed by using the quasi-static approximation [26] as

$$f_T = \left.\frac{g_m}{2\pi C_G}\right|_{V_D=V_{DD}} \qquad (1)$$

where $g_m$ is the transconductance, and $C_G$ is the gate capacitance computed as the derivative of the charge in the channel with respect to the gate voltage. As can be seen, MOSFET has ~30 % higher $f_T$ as compared to the SBFET counterpart. For what concerns instead the intrinsic switching time $\tau$, which represents the typical figure of merit for digital applications, we have used a previously developed comparison method that takes into account the power supply, on and off states [27]. This quantity is typically used to estimate the time it takes an inverter to switch, when its output drives another inverter. Figure 4b shows the intrinsic delay as a function of on-off ratio: in this case, MOSFET exhibits ~20 % faster switching speed than a middle-bandgap SBFET. The very high cutoff frequency and the very small delay shown in Fig. 4 are due to the extremely short channel length (15 nm) and the assumption of purely ballistic transport. In general, $f_T$ is inversely proportional to the channel length, and for longer channel SBFETs, for example, it can be expressed as $f_T \approx$ 73 GHz / ($L_{ch}$ in μm) at on state. In addition, additional parasitic capacitance could largely affect the estimated $f_T$ and delay.

*B. Atomistic Vacancy*

We now focus our attention on the effect of a single vacancy defect on device performance.

Figures 5a-b show the transfer characteristics for SBFET and MOSFET both in the linear and in the logarithmic scale, for different positions of a defect. All defects are placed in the middle of the channel along the width direction, whereas three different positions along the propagation direction are considered: in particular, the defect has been placed in correspondence of the source, the middle of the channel, and the drain.

As shown in Fig. 5, the defect near the source has the largest effect in both devices. As compared to the ideal device, the defect results in 46 % and 17 % smaller $I_{on}$ in SBFET and MOSFET, respectively. This is because the carrier transport in the device is totally controlled by the Schottky barrier at the source end for SBFETs, and by the top of the barrier, which is also located near the source, for MOSFETs.

The details of the $I_{on}$ reduction can be explained by the reduced quantum transmission and self-consistent electrostatic effect. For an SBFET with a defect near the source, thicker SB is induced (Fig. 6a) due to the electron accumulation, and quantum transmission is reduced (Fig. 6b) at the on state, which result in a smaller $I_{on}$. When a defect is located at halfway along the channel or near the drain of an SBFET, accumulated electrons lifts up potential barrier and reduces the energy window of electron injection from the source to channel, which results in reduced current with a lattice vacancy. In case of a MOSFET with a lattice vacancy near the source, self-consistent potential barrier is not increased, as shown in Fig. 6c. Instead, the reduced number of propagating states due to the lattice vacancy reduces the transmission probability (Fig. 6d), which results in a smaller on current. On the other hand, defects near the drain and in the middle of the channel do not affect device transfer characteristics as much as the case near the source. Transport is indeed mostly determined by the top-of-the-barrier potential, which, as shown in Fig. 6c, is only partially influenced by the presence of the defect in correspondence of the drain (and in the middle of the channel).

Next we show that the transfer characteristic is also very sensitive to the position along the channel width direction. The position of a defect varies from the center to the near-edge as shown in the inset of Fig. 7c. For both devices it has the largest effect on the $I_{on}$ when it is located at the position marked in-between. Because $N$=12 A-GNR has the largest effective coupling strength at that position [28], the device has severely reduced transmission (Figs. 7b and 7d) and hence the smallest on current (Figs. 7a and 7c). In comparison, it only has small effects when the defect is at the center or near the edge due to the relatively small effective coupling strength.

*C. Edge Roughness*

State-of-the-art patterning technique is far from atomic scale precision, and edge roughness of GNR is always expected in fabrication process. Therefore, it would be very useful to examine the effect of edge roughness on device performance. One of the simplest irregular-edge GNRs is depicted in Fig. 8a, which is obtained by removing carbon atoms from both edges in the same probability. In general, the off currents are increased due to the gap states induced in the band-gap region, which enhances the leakage current at the off state [22]. Figures 9a and 9c clearly show the local density of states in the band-gap region for an SBFET and a MOSFET, respectively, at off state. On the other hand, the on currents are generally decreased due to the reduced quantum transport [22]. Even though the gap states near the beginning of the channel may facilitate quantum transport, overall quantum transmission is reduced by the carrier transport through the imperfect-edge GNR as shown in Figs. 9b and 9d. For the structure of Fig. 8a, $I_{off}$ is increased by a factor of 7 and 4, and $I_{on}$ is reduced by 40 % and 20 % for an SBFET and a MOSFET, respectively.

In order to investigate the general behavior of GNRFETs with edge roughness, randomly generated 100 samples are simulated. Figure 10 is a histogram of $I_{on}$ for SBFETs in the presence of edge roughness, where carbon atoms are randomly added into or removed from the edges of GNR with probability $P$=0.05. The result shows that $I_{on}$ is generally decreased by edge roughness, and the mean value is 25 % smaller than the ideal one. In addition, the performance variation can be very large from device to device, which is caused by the different atomistic details of each irregular-edge GNR.

*D. Ionized Impurity*

The last non-ideality is an ionized impurity, which can exist near the GNR channel. In this study, Li ion is used as impurity, which has a positive 0.4q at 1.84 Å away from the GNR surface according to *ab-initio* calculations [29]. It is located in the middle of the GNR width at different positions along the transport direction. Figure 11 shows $I_D$-$V_G$ curves in the presence of an ionized impurity. For SBFETs, it has the largest effect with 20 % larger $I_{on}$ when located near the source because of the severely reduced Schottky barrier at source end (Fig. 12a), which is a key factor to determine the carrier transport in tunneling devices. If an impurity is located far from the source, the alteration of Schottky barrier is significantly reduced, and it only has a small effect on the $I_{on}$. On the other hand, an ionized impurity always has a relatively small effect on the $I_{on}$ of MOSFETs because it has a very limited influence over the barrier height. Regardless of the impurity position, the on current of MOSFET varies by less than 10 % (Fig. 11b).

Next, we simulated 100 cases at randomly distributed positions maintaining the distance between Li ion and GNR surface to explore its general effect on the $I_{on}$. Figure 13a is a histogram of $I_{on}$ for SBFETs in the presence of a positive ionized impurity, which shows two distinct groups. The first group has increased $I_{on}$ due to the severely reduced Schottky barrier when an ionized impurity is very close to source (0 < $x$ < 4 nm) or drain electrodes (13 < $x$ < 15 nm). Thirty four percent of the samples are counted in this group. On the other hand, $I_{on}$ of the second group is reduced, when an impurity is not located near the source or the drain, due to the quantum-mechanical reflection of non-uniform electrostatic potential. For what concerns MOSFET instead, the largest number of samples lay around the ideal value (12.5 µA), while the remaining samples differ by less than 8 %. Such insensitiveness is due to the fact that for the considered simulations the top of the barrier is well below the Fermi level of the source: local changes of the potential do not influence the overall source-to-drain current.

So far we focused on a positive ionized impurity near the GNR surface. In order to investigate the effect by a negative charge impurity, an external impurity of an electron is placed at 0.5 nm away from GNR surface. The electron is located in the middle of the GNR width at different positions along the transport direction. An electron impurity increases the self-consistent electrostatic potential, which is common for both SBFETs and MOSFETs, as shown in Fig. 14b and 14d. Therefore, the on current is decreased by 33 ~ 47 % in the presence of electron impurity. Even though MOSFETs are nearly invariant to a positive ionized impurity, they are very susceptible to a negative charge impurity.

## IV. CONCLUSION

In this study, GNR SBFETs and MOSFETs are compared by solving Schrödinger equation self-consistently with 3D Poisson equation. In ideal devices, MOSFETs show better device characteristics over SBFETs: larger maximum achievable on-off ratio, 50 % larger on current, larger transconductance, and better saturation behavior with 60 % smaller output conductance. Switching and high-frequency performance of GNR devices are also better in MOSFETs, which have 30 % higher cutoff frequency and 20 % faster switching speed.

Even under the influence of a defect or an impurity, MOSFETs are more robust than SBFETs. In the presence of a single lattice vacancy, $I_{on}$ of SBFET can be reduced by 46 %, which is much larger than that of MOSFET, due to severely affected Schottky barrier thickness of the tunneling device. Edge roughness of GNR can result in larger off current and smaller on current, in general, and the variability of device performance is very large because of the totally different atomistic configuration of GNR in such small channel devices. In the presence of a positive ionized impurity, $I_{on}$ of SBFET can be increased by 20 %, but its effect on MOSFET is very limited because the top of the barrier is nearly invariant to the positive impurity. However, a negative charge impurity always disturbs the carrier transport of GNRFETs due to the locally increased electrostatic potential.

**FIGURES**

Figure 1  *Simulated device structure* (a) Schottky barrier (SB) field-effect transistor (FET) with metal contacts (b) MOSFET with doped source and drain extensions. $SiO_2$ gate insulator is 1.5 nm thick with a relative dielectric constant $\kappa = 3.9$. $N=12$ armchair-edge graphene nanoribbon (GNR) is used as a channel material, which is 15 nm long, 1.35 nm wide, and the bandgap is $E_g \approx 0.6$ eV. The Schottky barrier height in Fig. 1a is a half band gap.

Figure 2  $I_D$-$V_G$ characteristics of (a) an ideal SBFET and (b) an ideal MOSFET. For a fair comparison between two different devices, the minimal leakage current $I_{min}$ of SBFET is chosen as a common off current $I_{off} = 10^{-7}$ A, and on state is defined at $V_{on} = V_{off} + V_{DD}$, where $V_{DD} = 0.5$ V is the power supply voltage. Gray windows in Figs. 2a and 2b show the operating voltage ranges of each device. (c) Transfer characteristics of the ideal devices after gate work function engineering, by which $V_{off}$ can be shifted to $V_G = 0$V. An ideal MOSFET has 50 % larger $I_{on}$ than an ideal SBFET. (d) $I_{on}$ vs. $I_{on}/I_{off}$. MOSFETs can have a significantly larger on-off ratio than SBFETs.

Figure 3  (a) $I_D$-$V_D$ characteristics at $V_G = V_{DD} = 0.5$ V. (b) Output conductance $g_d$ vs. $V_G$ for $V_D = V_{DD}$. MOSFET shows better saturation behavior, which can also be pointed out by smaller $g_d$.

Figure 4  (a) Cutoff frequency $f_T$ vs. $V_G$. (b) Intrinsic delay $\tau$ vs. $I_{on}/I_{off}$. MOSFETs can have higher

cutoff frequency and smaller intrinsic delay than SBFETs.

Figure 5    *The effect of a lattice vacancy along the transport direction $I_D$-$V_G$* of (a) SBFETs and (b) MOSFETs in the presence of a single lattice vacancy, in a log scale (left axis) and in a linear scale (right axis). The lattice vacancy is placed in the middle of the channel width direction, and at the different positions along the transport direction: near the source, in the middle of the channel, and near the drain.

Figure 6    Conduction band profile along the channel position for (a) SBFETs and (c) MOSFETs in the presence of a lattice vacancy at the on state. Energy-resolved current spectrum for (b) the SBFET and (d) the MOSFET in the presence of a vacancy near the source.

Figure 7    *The effect of a lattice vacancy along the channel width direction $I_D$-$V_G$* of (a) SBFETs and (c) MOSFETs in the presence of a single lattice vacancy. The vacancy is located at different positions along the width direction: near edge (solid line), at center (dash-dot line), and between the two (dashed line) as shown in the inset of Fig. 7c. The position of the defect along the transport direction is close to the source. Energy-resolved current spectrum for (b) the SBFETs and (d) the MOSFETs.

Figure 8    *The effect of edge roughness* (a) Atomistic configuration of a simulated GNR channel in the presence of edge roughness. $I_D$-$V_G$ characteristics of (b) the SBFET and (c) the MOSFET with the GNR channel shown in Fig. 8a.

Figure 9  Local density of states (LDOS) at the off state ($V_G = 0$ V, $V_D = V_{DD}$) for (a) the SBFET and (c) the MOSFET with the GNR channel of Fig. 8a. Energy-resolved current spectrum at the on state ($V_G = V_D = V_{DD}$) for (b) the SBFET and (d) the MOSFET. The solid lines in Figs. 9a and 9c show the band profiles of ideal transistors.

Figure 10 Histogram of $I_{on}$ for SBFETs in the presence of edge roughness of GNR by adding or removing carbon atoms with probability $P = 0.05$. One hundred samples are randomly generated and simulated. Mean is 6.36 µA, median is 6.31 µA, and standard deviation is 2 µA.

Figure 11 *The effect of a positive ionized impurity $I_D$-$V_G$* of (a) SBFETs and (b) MOSFETs in the presence of an ionized impurity, in a log scale (left axis) and in a linear scale (right axis). The impurity is located in the middle of the GNR width direction, and at the different positions along the transport direction. Li ion is used as impurity, which has +0.4q at 1.84 Å away from the GNR surface.

Figure 12 Conduction band profile along the channel position for (a) SBFETs and (c) MOSFETs in the presence of a positive ionized impurity at the on state. Energy-resolved current spectrum for (b) the SBFET and (d) the MOSFET in the presence of a positive ionized impurity near the source.

Figure 13 Histogram of $I_{on}$ for (a) SBFETs and (b) MOSFETs in the presence of an ionized impurity

with +0.4q at 1.84 Å away from GNR surface.

Figure 14 *The effect of a negative charge impurity $I_D$-$V_G$ characteristics of (a) SBFETs and (c) MOSFETs in the presence of a charge impurity with –q at 0.5 nm away from GNR surface. Conduction band profiles along the transport position at the on state for (b) the SBFETs and (d) the MOSFETs.*

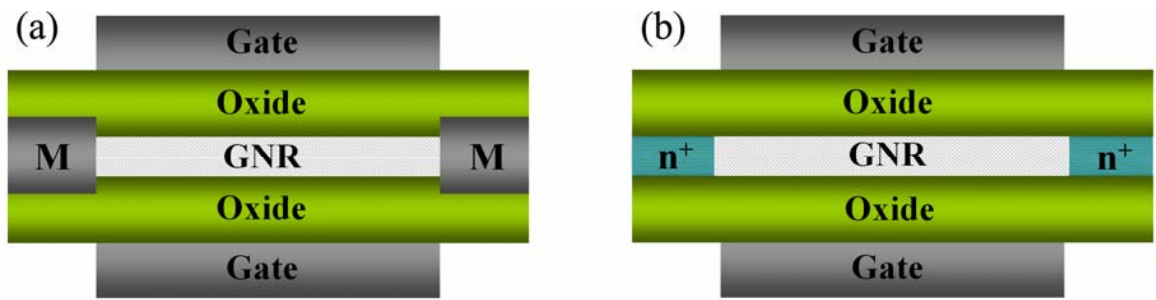

Figure 1

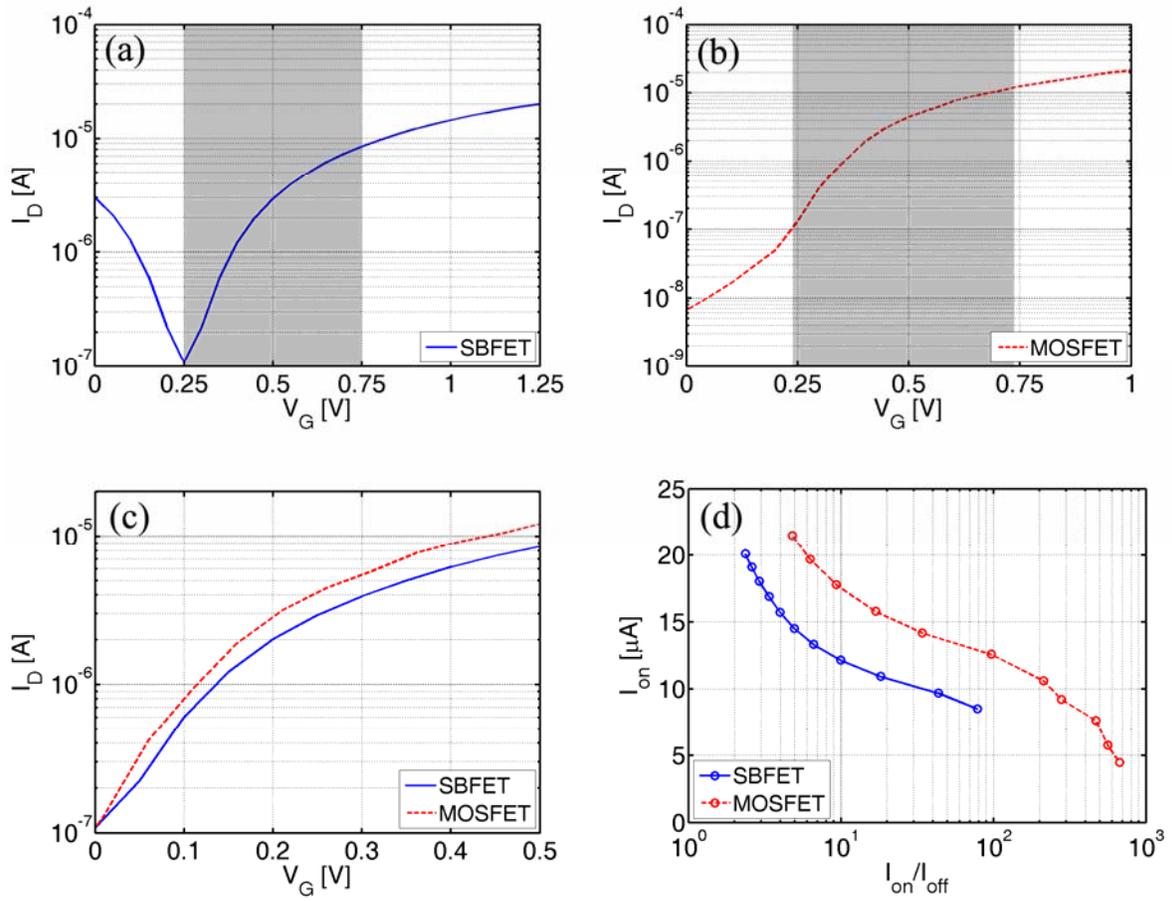

Figure 2

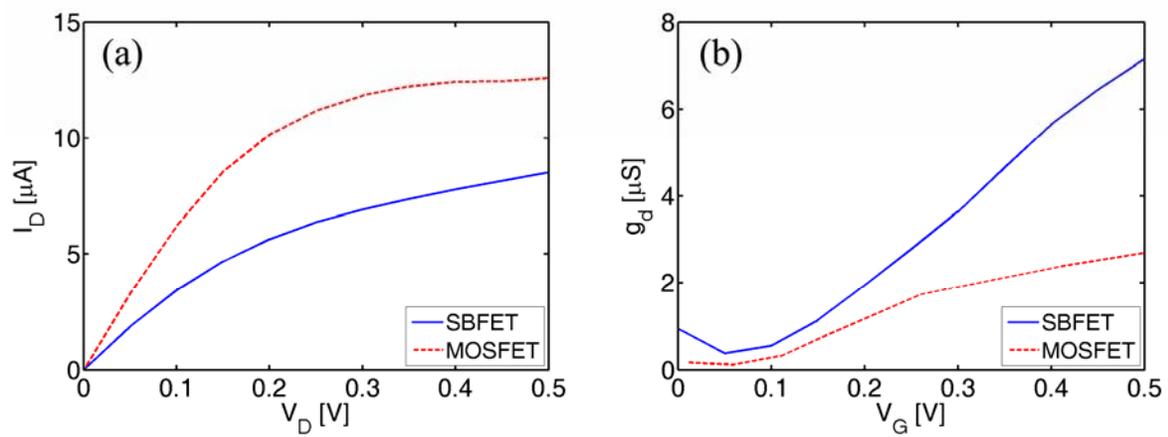

Figure 3

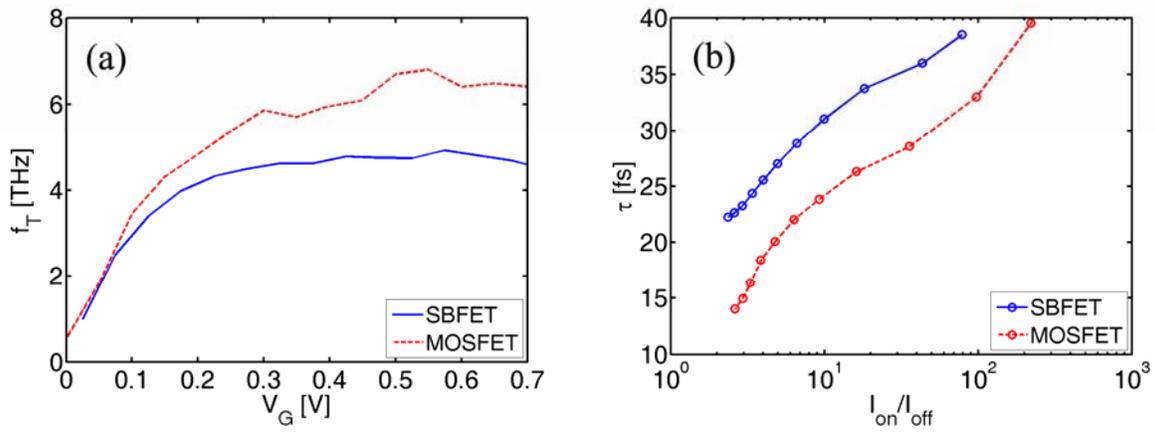

Figure 4

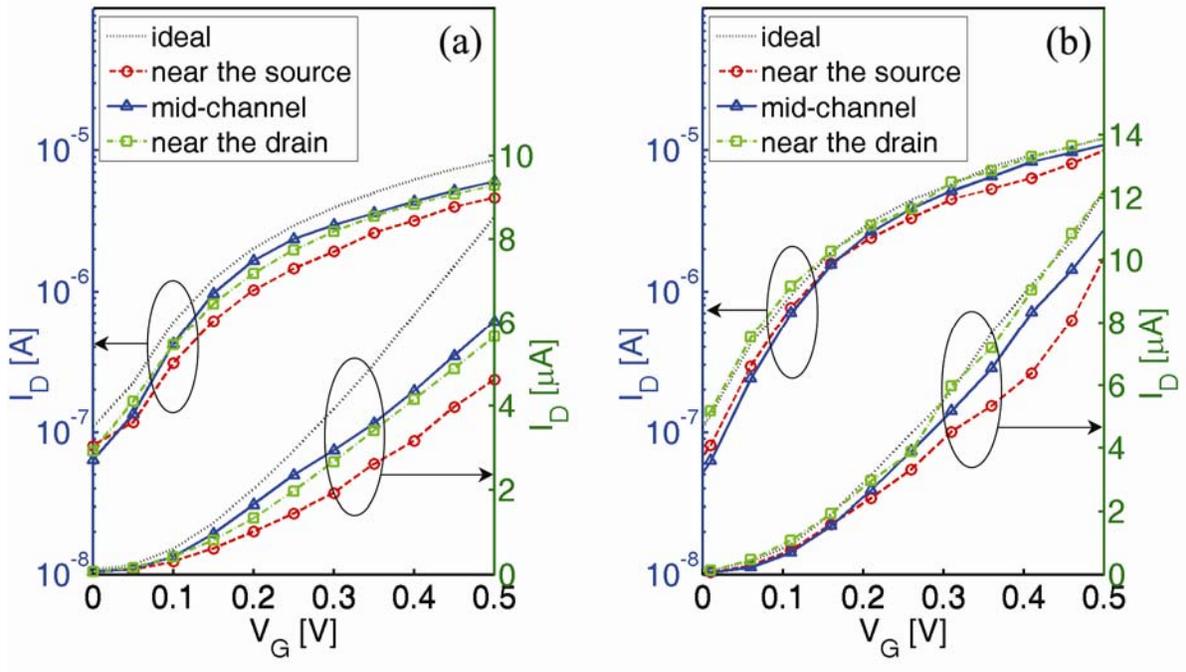

Figure 5

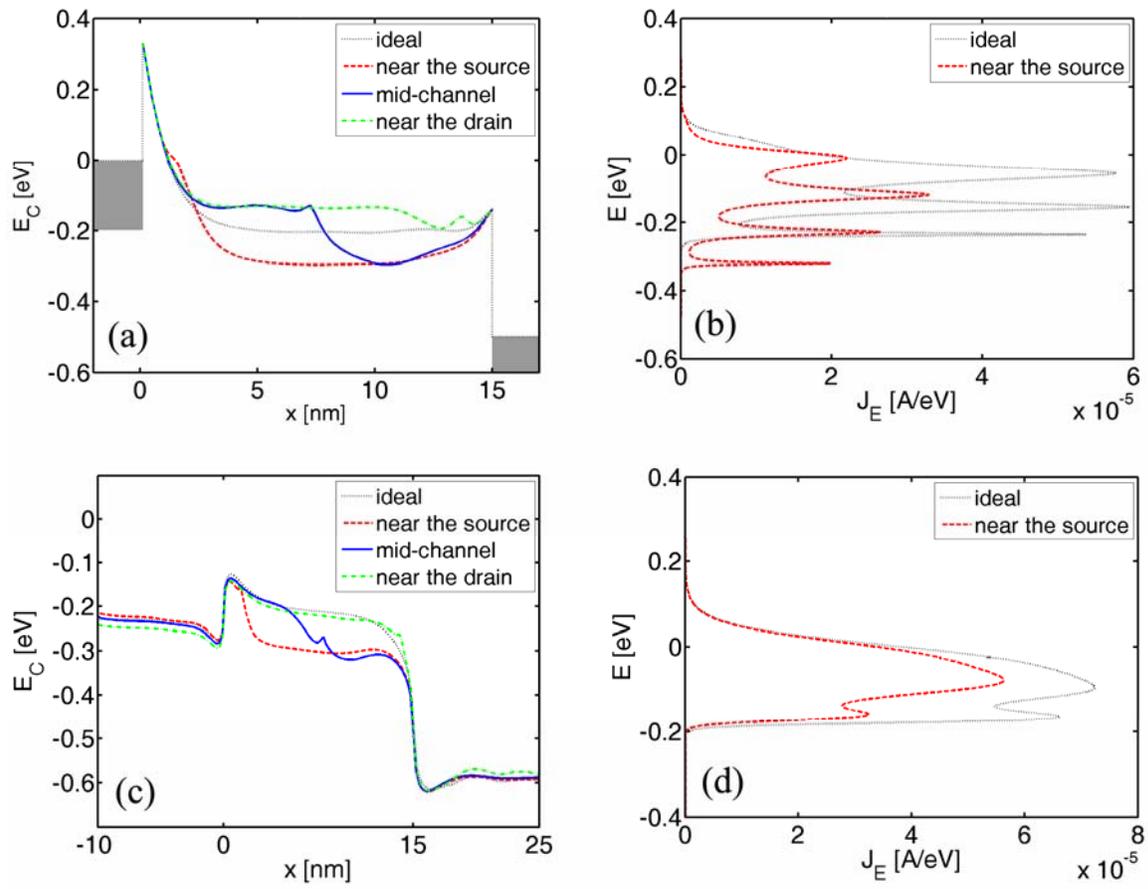

Figure 6

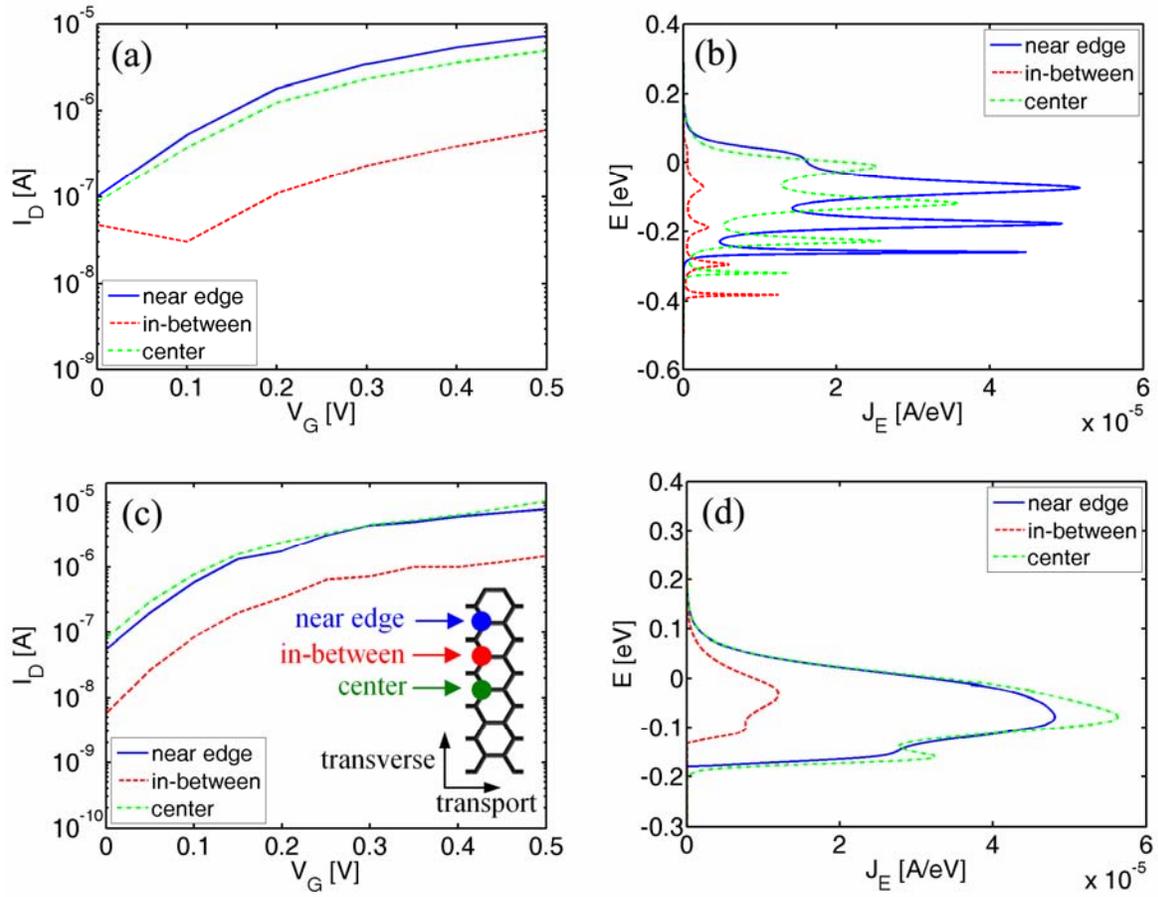

Figure 7

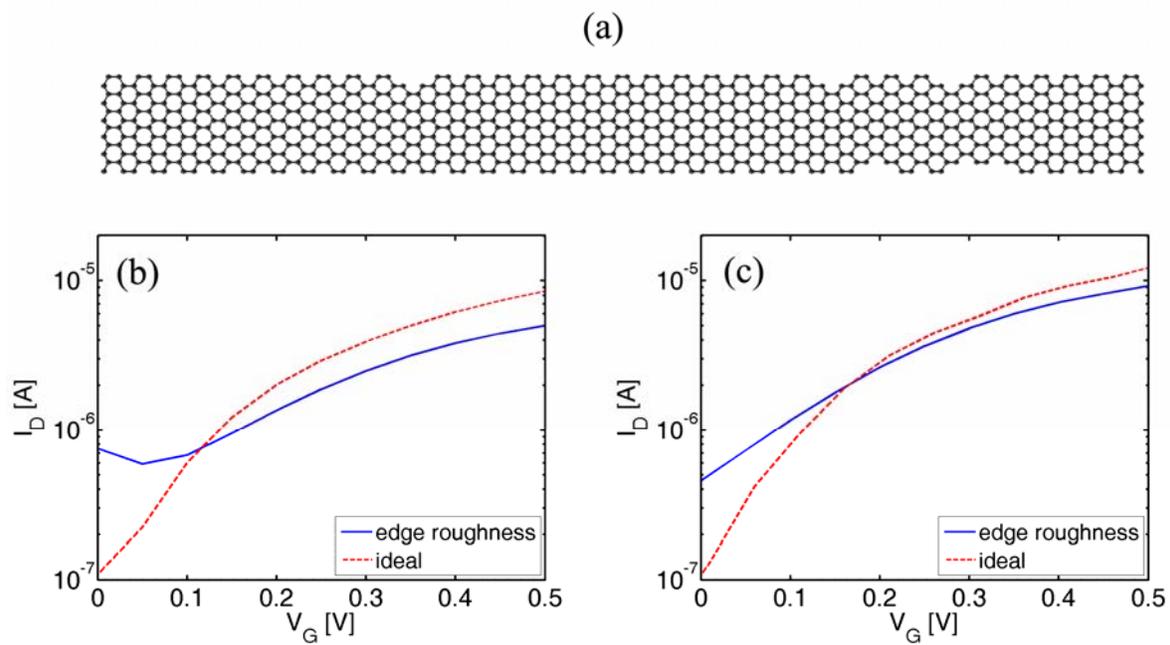

Figure 8

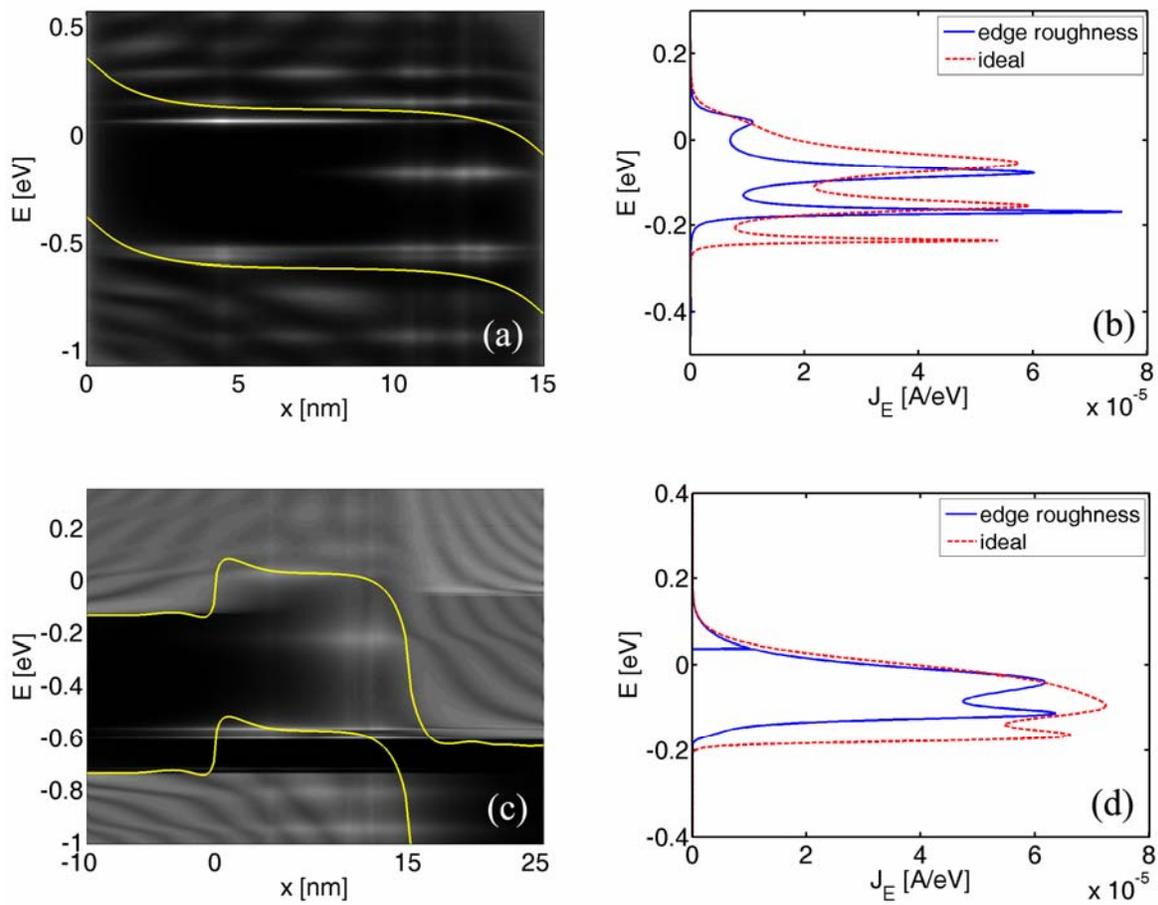

Figure 9

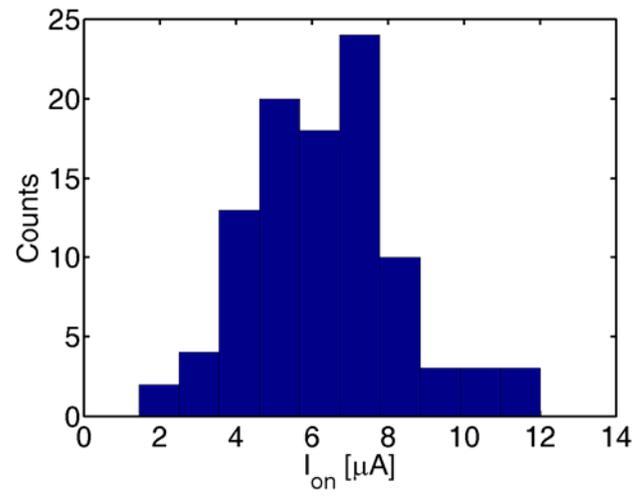

Figure 10

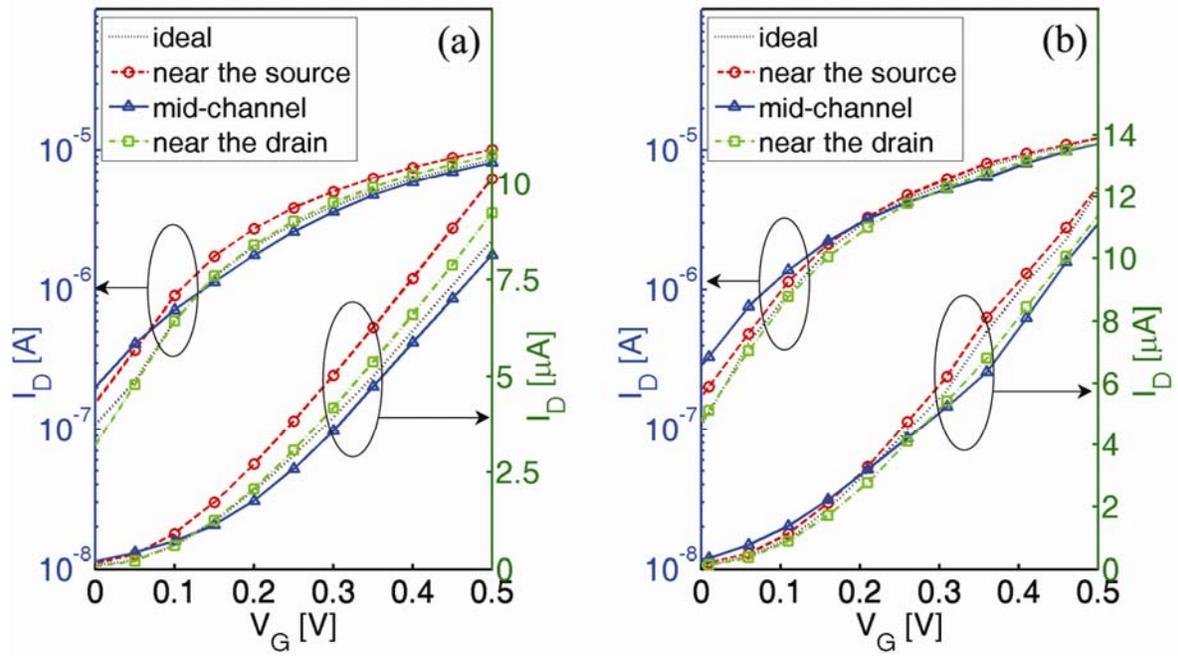

Figure 11

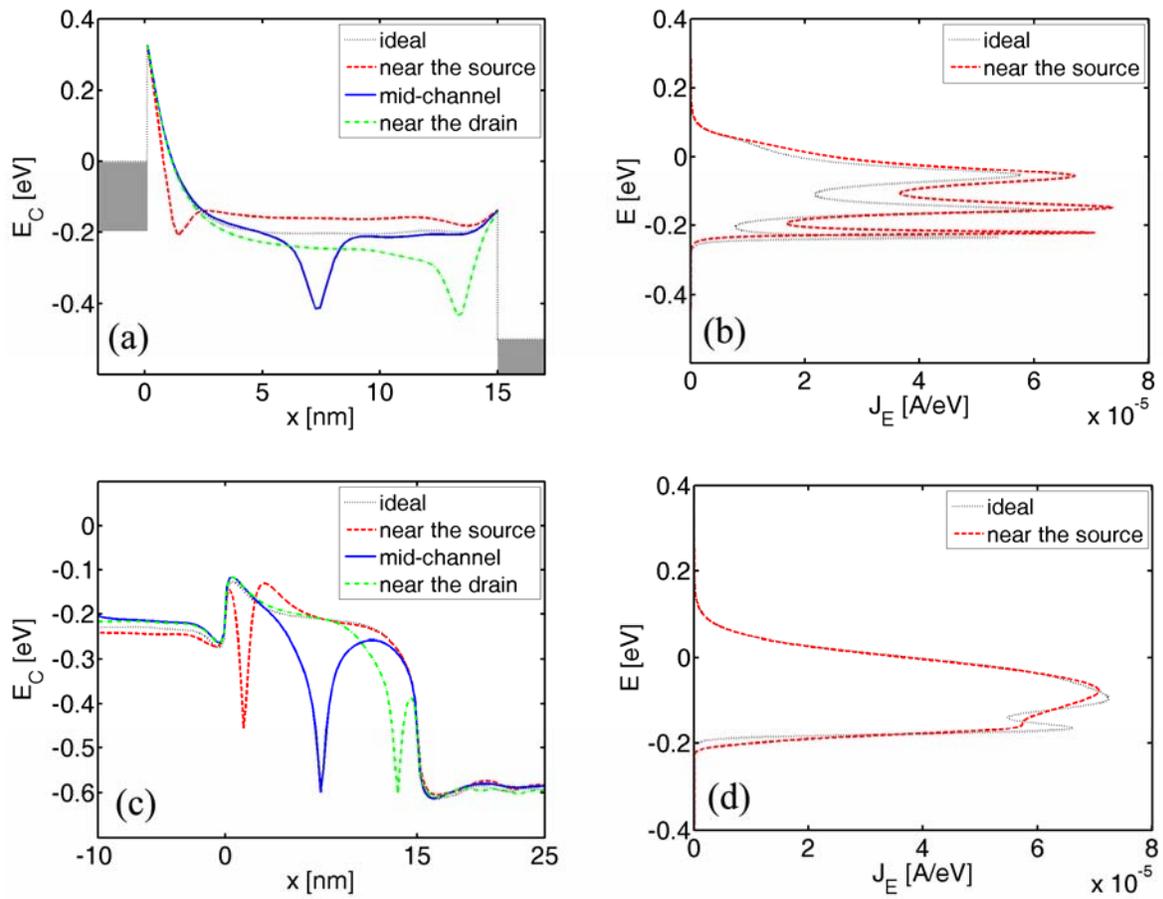

Figure 12

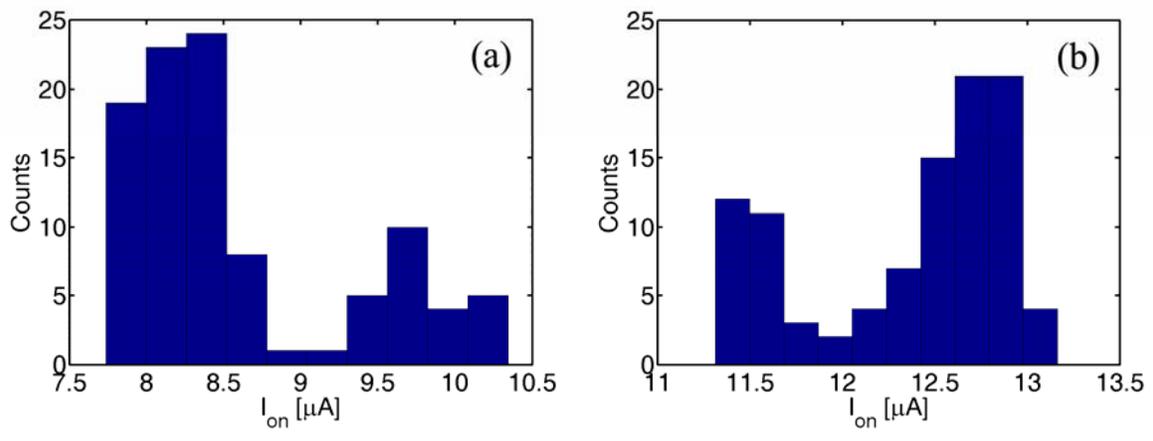

Figure 13

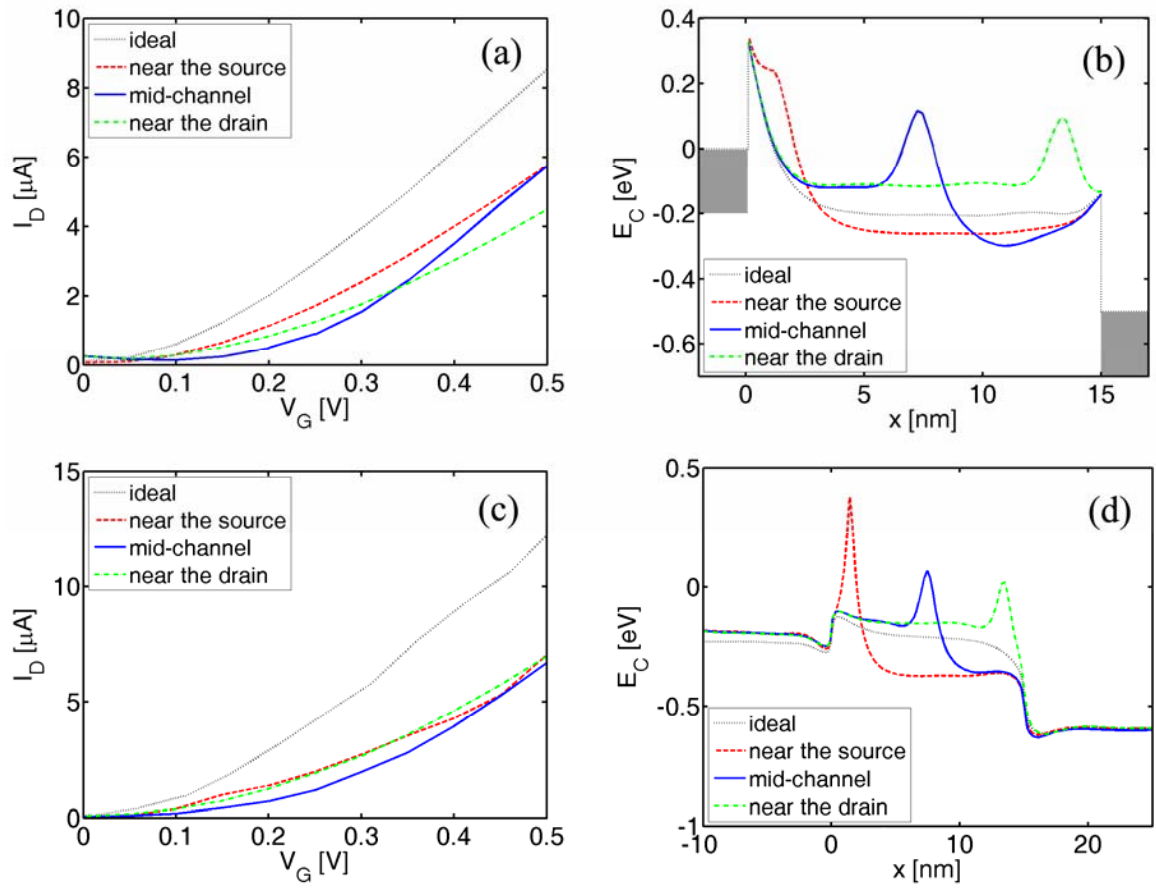

Figure 14